# Pressure-enhanced superconductivity and its correlation with suppressed resistance dip in (La,Pr)$_3$Ni$_2$O$_7$ films

Jinyu Zhao[1]*, Guangdi Zhou[2,3]*, Shu Cai[1]*, Shuaihang Sun[1]*, Yaqi Chen[3], Jing Guo[4], Yazhou Zhou[4], Haoliang Huang[2,3], Jin-Feng Jia[2,3,5], Yang Ding[1], Qi Wu[4], Zhuoyu Chen[2,3]†, Qi-Kun Xue[2,3,6]† and Liling Sun[1,4]†

[1]*Center for High Pressure Science & Technology Advanced Research, Beijing 100193, China*
[2]*Quantum Science Center of Guangdong-Hong Kong-Macao Greater Bay Area, Shenzhen 518045, China*
[3]*State Key Laboratory of Quantum Functional Materials, Department of Physics, and Guangdong Basic Research Center of Excellence for Quantum Science, Southern University of Science and Technology, Shenzhen, 518055, China*
[4]*Institute of Physics, Chinese Academy of Sciences, Beijing 100190, China*
[5]*State Key Laboratory of Micronano Engineering Science, Tsung-Dao Lee Institute & School of Physics and Astronomy, Key Laboratory of Artificial Structures and Quantum Control, Shanghai Jiao Tong University, Shanghai 200240, China*
[6]*Department of Physics, Tsinghua University, Beijing 100084, China*

The discovery of superconductivity with a transition temperature ($T_c$) exceeding 40 K in La$_3$Ni$_2$O$_7$ and (La,Pr)$_3$Ni$_2$O$_7$ thin films at ambient pressure provides a viable platform for the experiments that can only be conducted under ambient-pressure conditions, and for the theoretical investigations aimed at understanding the commonalities and peculiarities of the behaviors related to the superconductivity between the film and the compressed bulk systems - including the effects of oxygen vacancies and strain. Consequently, it is crucial to determine whether $T_c$ can be further enhanced and to uncover the underlying physics that controls the $T_c$ value in these ambient-pressure superconducting thin films. Here, we report a systematic study of hydrostatic pressure effects on the superconducting properties of (La,Pr)$_3$Ni$_2$O$_7$ thin films. We find that external pressure universally enhances $T_c$ of the film samples regardless of their initial $T_c$ value. The onset $T_c$ of 68.5 K at 2.0 GPa demonstrates a notable increase from 62 K at 0.3 GPa. Furthermore, we observe that the samples without zero resistance show a resistance dip just above the superconducting transition, whereas the samples that exhibit zero resistance do not display this dip. Applying pressure can suppress the dips and drive the system toward zero resistance. Based on our results, we propose that this feature is associated with oxygen vacancies and that the depth of the dip can serve as an indicator of the concentration of the vacancies. It is plausible that the dip is caused by the localization of mobile electrons at the vacancy sites. Applying pressure can delocalize these electrons, which in turn may contribute to the increase in $T_c$.

Since the discovery of high-temperature superconductivity (HTSC) in cuprates [1,2] and Fe-based superconductors [3], the search for a new family with high superconducting transition temperature ($T_c$) has been a central theme in condensed matter physics. This pursuit is driven by the goal of not only achieving higher $T_c$ but also deciphering the mechanisms of HTSC. Recently, Ruddlesden-Popper (RP) bilayer nickelates were found to exhibit signatures of superconductivity with $T_c$ near 80 K in the compressed $La_3Ni_2O_7$ bulk materials [4-6], and the superconductivity with $T_c$ about 30 K in the compressed $La_4Ni_3O_{10}$ single crystals [7]. Later, substantial efforts to enhance $T_c$ have been made, and $T_c$ has been raised to 96 K in $La_{1.57}Sm_{1.43}Ni_2O_{7-\delta}$ single crystals under pressure [8]. Recent breakthroughs in stabilizing superconductivity at ambient pressure through thin-film epitaxy growth have been achieved, demonstrating that specific strains arising from the interaction between the samples and appropriate substrates can induce ambient-pressure superconductivity in $La_3Ni_2O_7$, $(La,Pr)_3Ni_2O_7$ and $La_{3-x}Sr_xNi_2O_7$ thin films [9-15]. However, the $T_c$ observed in these thin films under ambient conditions remains lower than those found in bulk materials under high pressure. This gap suggests that further optimization is possible through combined tuning by both epitaxial strain and external pressure. The attempts to reduce this gap have made some progress recently. For example, applying hydrostatic pressure on $(La,Pr)_3Ni_2O_7/(La,Sr)_3Ni_2O_7$ films yielded an enhancement of onset $T_c$ up to ~ 61.5 K at 7 GPa [16], and employing an alternative growth method raises the ambient-pressure onset $T_c$ of $(La,Pr)_3Ni_2O_7$ thin films as high as 63 K [14].

Why is the superconducting transition temperature ($T_c$) of the thin films lower than the bulk counterparts? Recent high-pressure studies on $La_3Ni_2O_7$ bulk single crystals using the nitrogen-vacancy quantum sensors reveal that the superconductivity is driven by external pressure near 20 GPa with the shear stress below 2 GPa [17]. This suggests that hydrostatic pressure is expected to yield a higher $T_c$, because the electronic structure of nickelates is highly tunable and extremely sensitive to shear strain and structural distortions [13,16,18-24]. Moreover, the oxygen content can strongly influence the superconductivity of $La_3Ni_2O_7$ [25] - samples with oxygen content below 7 exhibit a reduced $T_c$. The same phenomenon is also observed in nickelate films [26].

One of the crucial questions is whether oxygen deficiency governs the normal-state conductivity and the resulting superconducting properties of the nickelates, and if yes, how pressure modifies the superconducting state in the systems with oxygen dediciency. In this high-pressure study, we focus our investigations on these issues.

First, we performed hydrostatic-pressure-resistance measurements on the $(La,Pr)_3Ni_2O_7$ films with various oxygen contents. As shown in Fig. 1a-1b, samples with relatively high oxygen content (S#1, S#2), characterized by a zero-resistance state at ambient pressure and their normal-state resistance exhibits typical metallic behavior ($dR/dT>0$) (see Supplementary Information-SI and Ref. 26). Under pressure, the normal-state resistance decreases monotonically at all temperatures, and the superconducting transition temperature ($T_c$) with zero resistance systematically shifts to higher temperatures. We also conducted high-pressure measurements on sample 3 (S#3) and observed similar high-pressure behavior: although S#3 did not exhibit zero resistance at ambient pressure, it approached near-zero resistance at around 2.0 GPa (Fig. 1c). The pressure-induced evolution of the superconducting transition is shown in Fig. 1d–1f for a clearer visualization. Specifically, for sample S#3, $T_c^{onset}$ increases significantly, reaching a maximum of ~ 68.5 K at 2.0 GPa (Fig. 1f). We determined the superconducting onset temperature ($T_c^{onset}$) by three methods: (1) the temperature at which the resistance deviates from the linear extrapolation of the normal state (as indicated by the black arrows in insets of Fig. 1e and 1f); (2) the jump in the $dR/dT$ derivative and (3) the two-line intersection temperature (see SI). It is seen that, for sample S#3 at 2.0 GPa, $T_c^{onset}$ is 68.5 K determined by method (1), 67.8 K by method (2), and 54.2 K by method (3). In order to remain consistent with earlier ambient-pressure studies of these bilayer superconducting nickelate [9,10,14], we employed the $T_c^{onset}$ from method (1) throughout the text.

Such enhancement is remarkable, especially in light of the moderate pressure range employed, suggesting that $(La,Pr)_3Ni_2O_7$ thin films are highly responsive to the hydrostatic compression. The pressure required to reach a 68.5 K superconducting transition is far lower than that needed to achieve a similar $T_c$ in bulk nickelates, highlighting that the substrate-imposed strain is more effective than external pressure in stabilizing superconductivity.

To establish whether pressure enhancement depends on the initial sample quality, we measured several films (S#4, S#5, and S#6) with different ambient-pressure $T_c^{onset}$ values. As shown in Fig. 2a-2c, although $T_c$s of these films are lower than those of S#1-S#3 and no zero resistance appears at ambient pressure due to oxygen deficiency [10,26], application of pressure can still enhance $T_c$ at a similar rate regardless of its initial transition temperature. The consistent behavior seen across compressed films indicates that pressure is an effective way to enhance the superconductivity of $(La,Pr)_3Ni_2O_7$ films.

In striking contrast to samples with high oxygen content, it is noted that those with relatively low oxygen content (S#4, S#5, and S#6) exhibited a more prominent feature in their resistance curves: there is a dip above the superconducting transition temperature (Fig. 2). This resistance dip is reminiscent of the behavior observed in under-oxidized nickelate samples, in which significant oxygen deficiency is present [26]. To illustrate this feature more clearly, we show the $R(T)$ data for these samples separately (Fig. 2d–2f). It is seen that the dip temperature ($T_{dip}$), marked by arrows, shifts to lower temperatures with increasing pressure, indicating that pressure can manipulate this resistance anomaly.

To quantify the level of the resistance dip, a signature of oxygen deficiency, we define its magnitude as $\Delta R = R_{Tc} - R_{Tdip}$ (see inset of Fig. 3a), where $R_{Tc}$ and $R_{Tdip}$ denote the resistance at onset of the superconducting transition and the resistance at the minimum, respectively. It is seen that the normalized dip magnitude, $\Delta R(P)/\Delta R(0)$, steadily decreases with increasing pressure in all samples (Fig. 3a), indicating that pressure can improve the superconducting properties through effectively healing the electronic structure featured by anomalous resistance behavior. We plotted the normalized dip temperatures ($T_{dip}(P)/T_{dip}(0)$) versus pressure for S#4-S#6 (Fig. 3b) and found that the data for all three samples show a similar trend that $T_{dip}$ decreases upon increasing pressure, revealing a similar improvement rate independent of their initial deficiency level.

Recent studies have highlighted the complex role of oxygen in nickelate superconductors [15,19,26-32]. The moderate oxygen annealing can improve superconductivity by filling oxygen vacancies [14,26], while excessive oxygen annealing can suppress or eliminate the superconducting state [29]. The prevailing understanding suggests that RP bilayers mainly possesses two distinct oxygen positions: oxygens within the Ni–O planes and apical oxygens of the octahedra [26-29]. Since pressure does not alter the sample's total oxygen content, along with the observations that the ambient-pressure $T_c$ generally correlates negatively with both oxygen deficiency [26] and the dip depth (Fig. 3a), we infer that the resistance dip observed here is associated with oxygen vacancies. These vacancies likely cause weak localization of mobile electrons, producing the resistance upturn prior to the superconducting transition. Applying pressure delocalizes these electrons, thus optimizing the superconducting electron state and increasing $T_c$ (Fig. 2, Fig. 3a and 3b). This hypothesis is supported by our pressure-release measurements: both $T_c$ and the associated resistance dip revert when the pressure is released (see SI), indicating that this process is associated with the elastic strain in the lattice.

Figure 3c shows the relationship between $T_{dip}$ and $T_c$ for the compressed S#4. The competition between two thermodynamic states is clearly seen in the pressure–temperature ($P$-$T$) color map of the resistance derivative ($dR/dT$): blue regions ($dR/dT > 0$) indicate metallic or superconducting state, while red regions ($dR/dT < 0$) mark the resistance upturn. As pressure increases, $T_{dip}$ is suppressed, while $T_c$ is enhanced. To verify whether this phenomenon is universal among samples with relatively low oxygen content, we plotted their $\Delta R(P)/\Delta R(0)$ versus $T_c(P)$-$T_c(0)$ in Fig. 3d. The data for all three samples fall on a common trend, implying that the suppression of oxygen vacancies is directly correlated with an increase in the superconducting transition temperature. These observations demonstrate that pressure suppresses the detrimental effect of oxygen vacancies, thereby favoring and stabilizing the superconducting ground state.

Finally, we summarize our results for all samples investigated in the $P$-$T$ phase diagram (Fig. 4a). Although the ambient-pressure $T_c$ values vary widely due to the difference in oxygen vacancies, the $T_c$s of all samples display a positive response - increasing linearly with increasing pressure. To demonstrate this consistent behavior, we show the increase in $T_c$ with pressure, $\Delta T_c$ ($P$), in Fig. 4b. Data from all samples follow a similar trend with $dT_c/dP$, demonstrating that the effects of the pressure-induced enhancement in $T_c$ are comparable across all the investigated bilayer nickelates and independent of the ambient-pressure oxygen content. Our results qualitatively explain the relation between the degree of oxygen vacancies and the depth of resistance dip, and strongly support the scenario that oxygen content plays a vital role in controlling the normal-state conductivity and the resulting superconductivity in the bilayer nickelate thin films.

In conclusion, we demonstrate that hydrostatic pressure can significantly enhance the superconductivity in $(La,Pr)_3Ni_2O_7$ thin films to a onset $T_c$ of 68.5 K under a moderate pressure of 2.0 GPa. The enhancement is universal across samples with varying initial $T_c$ values. Based on our experimental results and analysis, we propose that the observed resistance dip is related to the oxygen vacancies, which can serve as an experimental indicator for qualitatively evaluating the concentration of the oxygen vacancies. Our results show that this dip can be mitigated by applying pressure, indicating that oxygen vacancies, as a tunable control parameter for changing $T_c$, can be modulated by external pressure.

Thus, our findings highlight the critical importance of oxygen occupancy in determining the normal-state conductivity and the resulting superconductivity of bilayer nickelates, suggesting a practical pathway for further increasing $T_c$ of the films at ambient pressure, *i.e.* optimally engineering the strain state and tuning the oxygen content toward 7. It is expected that this work can provide both fundamental insights into the role of oxygen vacancies in stabilizing the superconductivity of nickelates and practical guidance for further materials optimization.

**Acknowledgements**

The work was supported by the National Key Research and Development Program of China (Grants No. 2021YFA1401800, No. 2022YFA1403900, No. 2024YFA1408101, and No. 2022YFA1403101), the National Natural Science Foundation of China (Grants No. No. 12474054, No. 12122414, No. 12274207, No. 92565303, No. 12374455, No. 12504166, No. 92265112, and No. 52388201), the Guangdong Major Project of Basic Research (No. 2025B0303000004), the Quantum Science Strategic Initiative of Guangdong Province (No. GDZX2501001, No. GDZX2401004, and No. GDZX2201001), the Municipal Funding Co-Construction Program of Shenzhen (No. SZZX2401001 and No. SZZX2301004), the Science and Technology Program of Shenzhen (No. KQTD20240729102026004). Partial work was supported by the Synergetic Extreme Condition User Facility (SECUF, https://cstr.cn/31123.02.SECUF). We acknowledge the support from the Station of Quantum Materials.

These authors with stars (*) contributed equally to this work.

Correspondence and requests for materials should be addressed to: Zhouyu Chen (chenzhuoyu@sustech.edu.cn), Qi-Kun Xue (xueqk@sustech.edu.cn) and Liling Sun (liling.sun@hpstar.ac.cn or llsun@iphy.ac.cn).

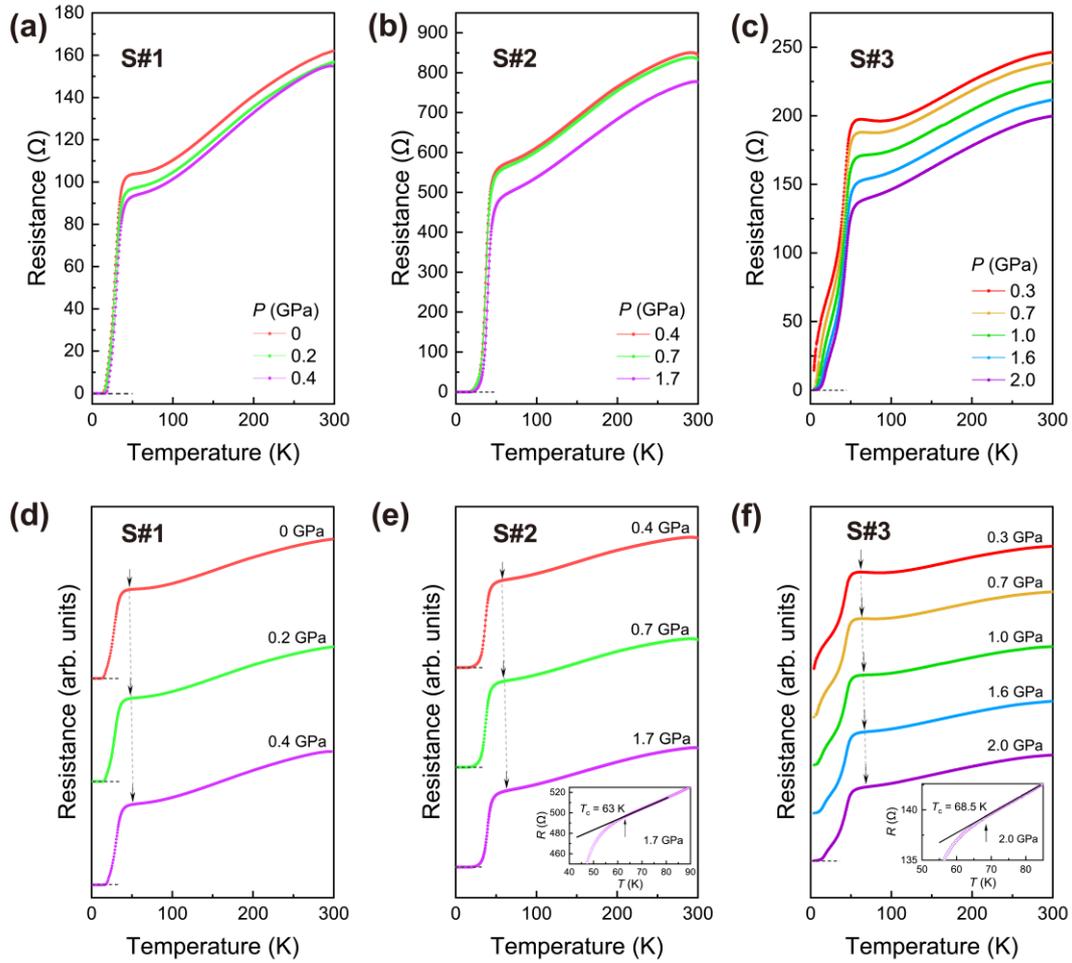

**Fig. 1. Transport properties of (La,Pr)$_3$Ni$_2$O$_7$ thin films with relatively high oxygen content at different pressures.** Temperature dependence of resistance for samples S#1 (**a**), S#2 (**b**), and S#3 (**c**). S#1 and S#2 exhibit superconducting transitions with a zero-resistance state for pressures ranging from 1 bar to 0.4 GPa (S#1) and to 1.7 GPa (S#2). Although S#3 does not attain zero resistance under low pressure, it approaches zero at 2.0 GPa. Normal-state resistance decreases with increasing pressure for all samples. Independent $R(T)$ for S#1 (**d**), S#2 (**e**), and S#3 (**f**) at different pressures to illustrate the evolution of the superconducting transition. The black arrows indicate the superconducting onset temperature $T_c^{onset}$. The insets in (**e**) and (**f**) demonstrate how $T_c^{onset}$ is determined for S#2 at 1.7 GPa and S#3 at 2.0 GPa - it is taken as the point where the resistance curve deviates from the linear fit to the normal-state resistance (black lines). The determination of $T_c^{onset}$ for S#1-S#3 at different pressures can be found in SI.

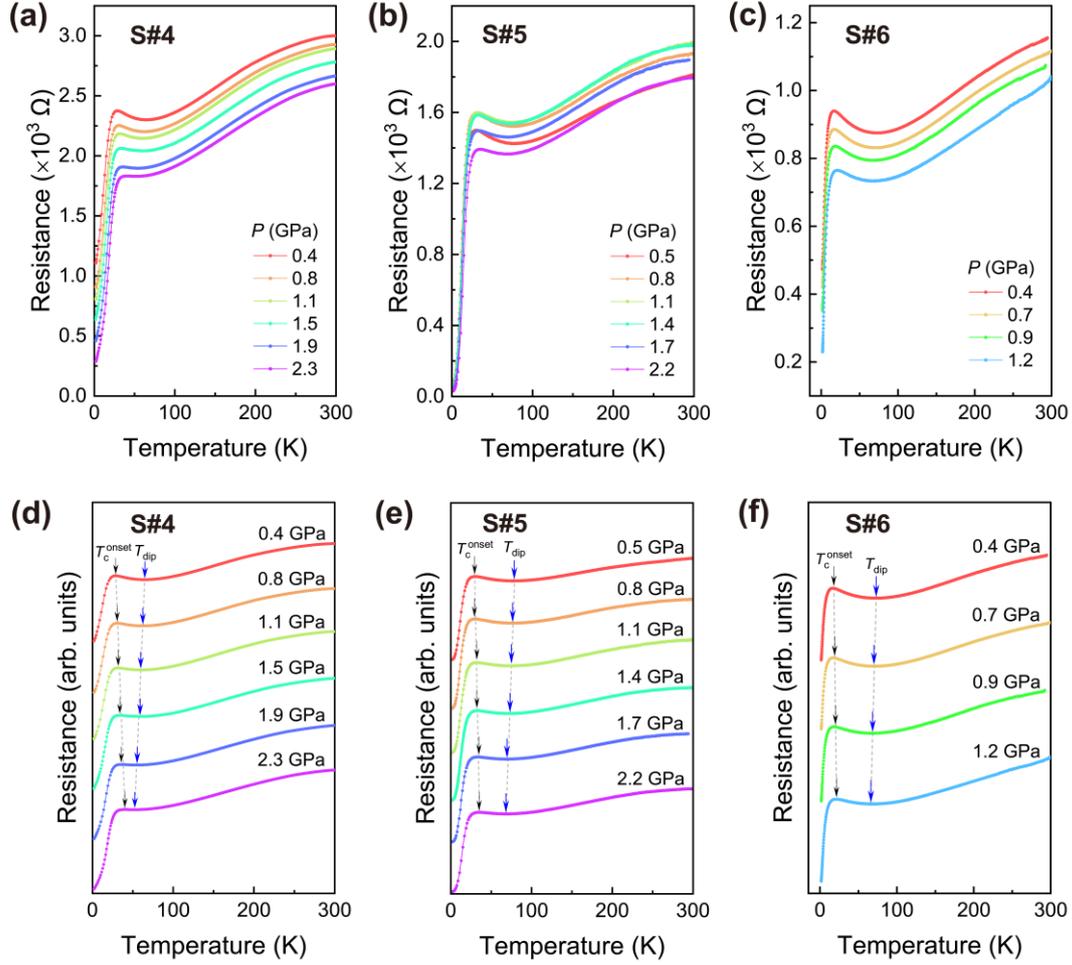

**Fig. 2. Transport properties of (La,Pr)$_3$Ni$_2$O$_7$ thin films with relatively low oxygen content at different pressures.** Temperature dependence of resistance for compressed S#4 (**a**), S#5 (**b**), and S#6 (**c**). Independent $R(T)$ data for S#4 (**d**), S#5 (**e**), and S#6 (**f**) for clarity. An obvious resistance dip can be observed before the superconducting transition. The black and blue arrows represent the onset temperatures of superconducting transition ($T_c^{onset}$) and the temperature of resistance dip ($T_{dip}$), respectively. The determination of $T_c^{onset}$ for S#4-S#6 at different pressures can be found in SI.

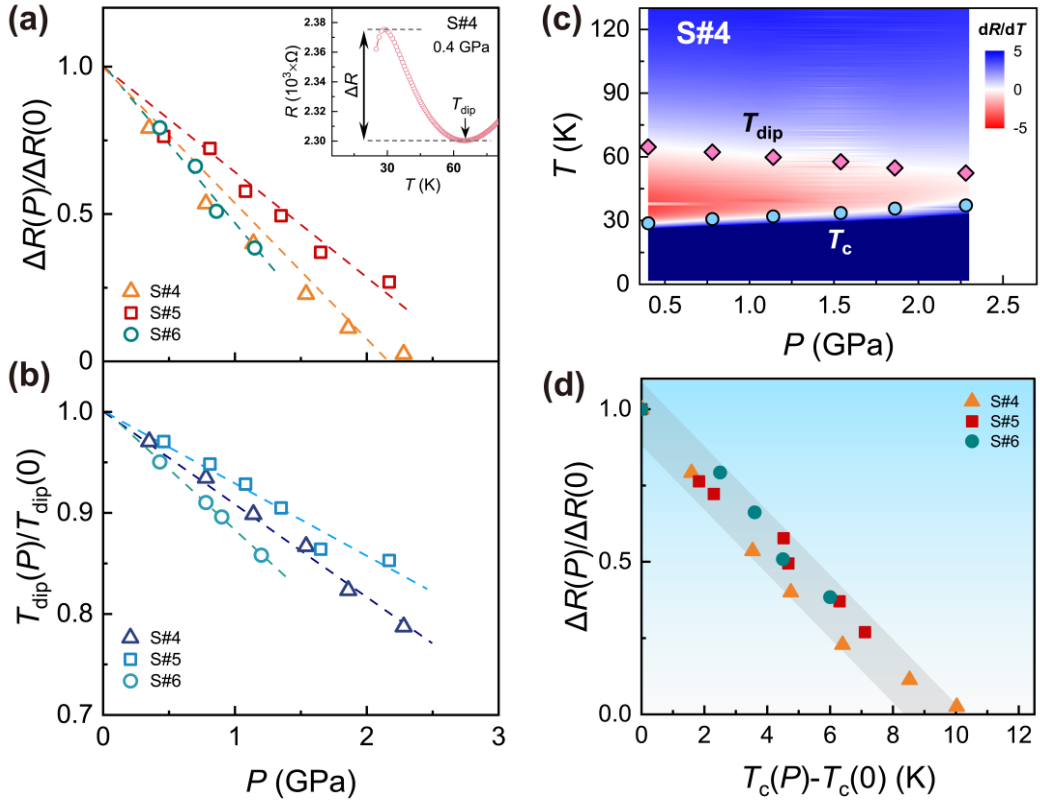

**Fig. 3. Pressure-induced universal suppression of the resistance dip and the dip temperature, as well as their correlations with superconductivity. a,** Normalized resistance dip magnitude, $\Delta R(P)/\Delta R(0)$, as a function of pressure for S#4, S#5 and S#6. The inset illustrates the definition of the dip magnitude ($\Delta R$) as the difference between the resistance at the onset of the superconducting transition and the resistance at the observed minimum. Dashed lines are guides to the eye, emphasizing the steady suppression of the resistance dip with increasing pressure. **b,** Pressure dependence of the normalized dip temperature $T_{dip}(P)/T_{dip}(0)$ for S#4, S#5, and S#6. The observed results reveal a universal suppression of the dip temperature across different samples. Dashed lines are drawn as guides to the eye. **c,** Pressure-temperature (*P-T*) color map for S#4, derived from the first derivative of resistance (d$R$/d$T$). The blue regions denote metallic behavior (d$R$/d$T >$ 0) and red regions stand for the resistance upturn (d$R$/d$T <$ 0). Diamonds and circles represent the temperatures of resistance dip ($T_{dip}$) and the superconducting transition ($T_c$), respectively, demonstrating that applied pressure suppresses the resistance dip and enhances superconductivity. **d,** Plot of $\Delta R(P)/\Delta R(0)$

versus $T_c(P)$-$T_c(0)$ for S#4-S#6. The shaded band is a guide to the eye.

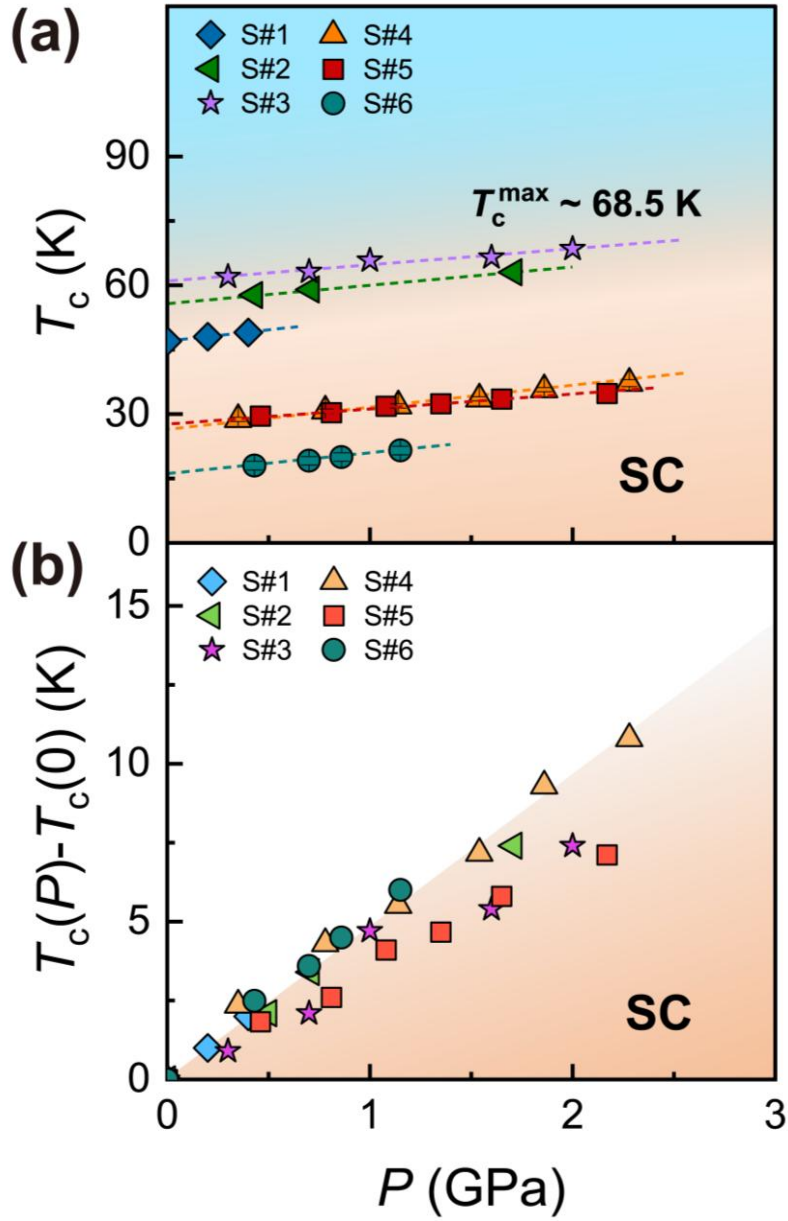

**Fig. 4. Superconducting phase diagram under hydrostatic pressure. a**, Pressure dependence of the superconducting onset temperature $T_c^{onset}$ for samples S#1–S#6. The variation in ambient-pressure $T_c^{onset}$ reflects the different oxygen contents across the samples. Dashed lines represent linear fits to the data, regardless of their initial $T_c$ values. The maximum $T_c^{onset}$ of approximately 68.5 K is observed from the compressed S#3 at 2.0 GPa. **b**, The pressure-induced enhancement, $T_c(P)$-$T_c(0)$, as a function of pressure. The data exhibits a consistent increase in $T_c$ with comparable linear slopes.